# PSTM Transaction Scheduler Verification Based on CSP and Testing

Miroslav Popovic

University of Novi Sad, Faculty of Technical Sciences, Trg Dositeja Obradovica 6, Novi Sad, Serbia, miroslav.popovic@rt-rk.uns.ac.rs

Marko Popovic

University of Novi Sad, Faculty of Technical Sciences, Trg Dositeja Obradovica 6, Novi Sad, Serbia, marko.popovic@rt-rk.uns.ac.rs

Branislav Kordic

University of Novi Sad, Faculty of Technical Sciences, Trg Dositeja Obradovica 6, Novi Sad, Serbia, marko.popovic@rt-rk.uns.ac.rs

Huibiao Zhu

East China Normal University, Shanghai Key Laboratory of Trustworthy Computing, Shanghai 200062, China, hbzhu@sei.ecnu.edu.cn

Many online transaction scheduler architectures and algorithms for various software transactional memories have been designed in order to maintain good system performance even for high concurrency workloads. Most of these algorithms were directly implemented in a target programming language, and experimentally evaluated, without theoretical proofs of correctness and analysis of their performance. Only a small number of these algorithms were modeled using formal methods, such as process algebra CSP, in order to verify that they satisfy properties such as deadlock-freeness and starvation-freeness. However, as this paper shows, using solely formal methods has its disadvantages, too. In this paper, we first analyze the previous CSP model of PSTM transaction scheduler by comparing the model checker PAT results with the manually derived expected results, for the given test workloads. Next, according to the results of this analysis, we correct and extend the CSP model. Finally, based on PAT results for the new CSP model, we analyze the performance of PSTM online transaction scheduling algorithms from the perspective of makespan, number of aborts, and throughput. Based on our findings, we may conclude that for the complete formal verification of trustworthy software, both formal verification and it's testing must be jointly used.

# 1 INTRODUCTION

Transactional memory (TM) was invented by Herlihy and Moss, as an extension of a cache-coherence protocol, in order to support transactions executed on multicores, which are operating on shared memory variables (called t-variables, briefly t-vars) [1-2]. Software TM (STM) was introduced as a software implementation of the TM abstraction [3]. STM transactions are easy to program (because they atomically execute simple sequential code), they cannot deadlock (because the fastest transaction gets committed whereas the other concurrent transactions get aborted and re-executed), and they provide great performance when contention among transactions is lower (because they are executed speculatively, i.e. without the overhead incurred by locks).

However, in case of high contention among transactions, the overall system performance may be affected, because many transactions may be aborted and re-executed, even worst some of them may be starved. Therefore, various transaction scheduler (also known as contention manager) architectures and algorithms were introduced in order to provide good performance even for high concurrency workloads, e.g. [4-6].

Python STM (PSTM) [7] is a general purpose STM for Python, which may be used in a wide range of application specific domains, from simulations in computational chemistry [8-9], to data science, to IoT based systems such as smart homes, vehicles, and cities. PSTM was successfully formally verified using three complementary approaches (see the section 1.1).

PSTM transaction scheduler architecture and the four online scheduling algorithms, named Round Robin (RR), Execution Time Load Balancing (ETLB), Avoid Conflicts (AC), and Advanced Avoid Conflicts (AAC), were developed with the main goal to minimize the makespan and consequently to maximize the throughput [10-11]. Informally, *makespan* for a given group of transactions is their total execution time (more formal definition is given at the beginning of the section 3.1). PSTM online transaction scheduling algorithms were experimentally evaluated and their time complexity bounds were proved in [11]. The first attempt to formally verify the algorithms RR, ETLB, and AC in [12] was made with some shortcomings, and the main goal of this paper is to remedy and complete this verification.

In this paper, we first analyze the previous CSP model of PSTM transaction scheduler introduced in [12] by comparing the model checker PAT results with the manually derived expected results, for the test workloads defined in [12]. Next, according to the results of this analysis, we correct and extend the CSP model. Finally, based on PAT results for the new CSP model, we analyze the performance of PSTM transaction scheduling algorithms from the perspective of makespan, speedup, number of aborts, and throughput.

From the perspective of engineering of computer based systems, this paper shows that using solely formal methods has its disadvantages by providing examples of possible shortcomings. This paper also offers an approach to discover these shortcomings by testing the model checker results, which is conducted by comparing the model checker results with the manually derived expected results. Therefore, as demonstrated in this paper, both formal verification and it's testing should be jointly used in order to conduct the complete formal verification of trustworthy systems.

The rest of the paper is organized as follows. The section 1.1 presents closely related work, the section 2 presents the analysis of the previous CSP model, the section 3 presents the new CSP model, the section 4 presents the performance analysis of the four PSTM online transaction scheduling algorithms, and the section 5 presents the paper conclusions.

## 1.1 Related Work

This section presents a brief overview of the most closely related research that was conducted before this paper.

*Formal verification* process used in this paper is based on model checking. *Model checking* is a technique for automatic verification of software and reactive system, and it consists in verifying some properties of the model of the system [22]. On the other hand, *testing* as defined in Cambridge dictionary is the process of using or trying something to see if it works, is suitable, obeys the rules, etc. [23]. In this paper, we are *testing* the formal verification process itself, in order to cross check whether it produces the expected theoretical results. To the best of our knowledge, this is the first paper that advocates and demonstrates such an approach to complete formal verification. So, testing used in this paper should not be confused with *software testing*, which is considered as woefully inadequate for detecting errors in highly concurrent designs [22].

A PSTM [7] based system is a typical client-server architecture, where PSTM is a server and transactions are its clients. Transactions request services from PSTM by calling the PSTM API functions. PSTM serves the requests by maintaining the system dictionary of shared t-variables. A t-variable is a tuple (*key*, *ver*, *val*), where *key*, *ver*, and *val* are t-variable key, version, and value, respectively. The two main PSTM API functions are getVars and commitVars.

A PSTM transaction gets some t-variables (by calling the API function getVars) and stores them in its local t-variable copies, does required processing (including updating some local t-variable copies), and finally commits all its operations on shared t-variables (by calling the API function commitVars). PSTM serves all the transaction requests atomically, one by one. That is the key idea behind the PSTM based architecture.

Since PSTM may be used as a component of a critical infrastructure, it was formally verified using three complementary approaches.

The first approach is based on the formal method Timed Automata (TA) [20] and its accompanying model checker UPPAAL [21]. The formal model of a PSTM based system [13] is constructed as a network of timed automata comprising automata representing: linear and cyclic transactions, the queue used by the remote procure call mechanism, and the transactional memory itself. The formal model was automatically analyzed by UPPAAL, and the following three properties were proved: safety (i.e. atomicity), liveness (one of the concurrent transactions must get committed), and termination (all the cyclic transactions must eventually be completed).

The second approach in [14] is based on the process algebra Communication Sequential Processes (CSP) [15], and the accompanying model checker PAT [16-17]. The formal model of a PSTM based system is hierarchically decomposed into two models. The higher level model is constructed as group of processes comprising an application process and a transaction process, whereas the lower level model is constructed as a group of processes comprising a transaction process, PSTM API process, PSTM server process, and PSTM dictionary process. The formal model was automatically analyzed by PAT, and the following three properties were proved: deadlock freeness, ACI (atomicity, consistency, and isolation), and optimism (essentially the same as the liveness proved for TA model).

The third approach is based on the push/pull semantic model [17]. The formal PSTM push/pull semantic model is constructed as the mapping of operations within the PSTM's generic transactional algorithm to the four relevant push/pull rules: PULL, APPLY, PUSH, and CMT [18]. The formal model was manually analyzed, and by proving that the model satisfies correctness criteria for the relevant push/pull rules, it was proved that PSTM satisfies serializability (i.e. sequential consistency).

The four PSTM online transaction scheduling algorithms (RR, ETLB, AC, and AAC) were developed hierarchically, from the simplest RR to the most advanced AAC, and they were compared from the perspectives of time complexity, quality of theoretical initial schedules, and experimentally measured speedup over RR and number of aborts [11]. As we go from RR, over ETLB and AC, to AAC, the algorithm's time complexity increases from $\Theta(1)$, over $\Theta(n)$ and $O(n \cdot m^2)$, to $O(n^2 \cdot m^2)$, where $n$ is the number of workers and $m$ is the number of t-variables used by transactions.

Theoretical schedules in [11] were manually derived for the three test workloads (CFW, RDW, and WDW), which were also used for experimental evaluation. As expected, only AAC makes optimal schedules, whereas other algorithms make suboptimal or even schedules with conflicts. The experimental results for the speedup over RR and the number of aborts, for the complete test workload execution, are well aligned with the theoretical results. This fact validates both the theoretical analysis and the algorithms' implementations in Python, as well as their experimental evaluation.

Generally, comparing results of independent system development phases is an important concept in engineering. Another important concept used in [11] is the analysis of theoretical schedules, and it will be also used in this paper in order to test the model checker results.

PSTM scheduler architecture and the first three online transaction scheduling algorithms (RR, ETLB, and AC) from [10] were formalized using the process algebra CSP in [12]. The formal model is constructed as a group of processes comprising an application process, the scheduler input queue process, the scheduler process, the worker input queue processes, the worker processes, and the processes formalizing the behavior of RR, ETLB, and AC algorithms.

The formal modelling process in [12], that is, the process of generating the formal model from the Python code (from [10]) was conducted manually in two phases. In the first phase: (i) Python FIFO queues (from the Python module multiprocessing): assigned to the scheduler process and to the *i*-th worker process were modelled as the CSP processes *QueueIn* and *Queue$^i$*, respectively, and (ii) the Python functions executed by the scheduler process and the i-th worker process were manually translated into the CSP processes *Scheduler* and *Worker$^i$*, respectively. In the second phase, CSP processes were manually translated into the CSP dialect called CSP#, which is recognized by the model checker PAT.

Unfortunately, in both phases some oversights took place, and they were not discovered, because the results of these steps were not checked against some expected results. Python code in [10] was tested using traditional software testing, but CSP and CSP# code in [12] was not tested. Unaware of the real situation, authors proceeded to proving desired properties by PAT.

The following two properties of the formal model were automatically proved by PAT: deadlock freeness and starvation freeness. Additionally, the performance of the processes representing transaction scheduling algorithms was automatically analyzed (by using the "with" option of the liveness properties encoded as "reaches" types of assertions) from the perspective of makespan, speedup, the number of aborts, and throughput.

So, the main contribution of [12] is that it provided a good foundation for future research, including the initial CSP model encoded in a PAT's dialect of CSP called CSP#, the selection and definition of correctness properties, and the selection of transaction scheduling algorithms' performance matrices. However, the formal verification in [12] was made with some shortcomings, as will be shown in the next section of the paper.

## 2 ANALYSIS

This section presents the analysis of the previous research conducted in [12]. The main goal of this analysis was to check whether the results of formal verification in [12] are aligned with the theoretical and experimental results in [11]. Perhaps surprisingly, we discovered that they were not. The next three subsections present the analysis method, theoretical schedules for test workloads in [12], and the analysis findings, respectively.

### 2.1 Analysis Method

The analysis method is based on analysis of theoretical schedules, which are expected to be produced by the subject online transaction scheduling algorithms for the given test workloads. The method comprises the following steps:

1. Derive theoretical schedules.
2. Calculate the makespan and the number of aborts.
3. Compare the results of the previous step with the results in [12].

One of the shortcomings of [12] is that instead of using the same test workloads that were previously used in [10-11], it introduced different test workloads named CFW, CW-1, and CW-2. All these test workloads are modeled in CSP# as arrays of five transactions [$T_0$, …, $T_4$] where each transaction operates on a single t-variable from the set of t-variables {$A$, $B$, $C$, $D$, $E$, $F$}. The duration of $T_0$ is 50 whereas the durations of $T_1$ to $T_4$ are 10 each (later in this paper we will assume that these unnamed time units are milliseconds). The usage of t-variables by a test workload is defined by an array of t-variables used by respective transactions. These arrays for CFW, CW-1, and CW-2 are [$A$, $B$, $C$, $D$, $E$, $F$], [$A$, $A$, $B$, $B$, $C$], and [$A$, $A$, $A$, $A$, $A$], respectively.

Since the new test workloads introduced in [12] are different from the test workloads previously used in [10-11] (the latter use more transactions, more t-variables per transaction, and more diverse transaction durations), direct comparison of results from [11] and [12] was not possible. Therefore, we derived the theoretical schedules for CFW, CW-1, and CW-2, which are presented in the next section.

### 2.2 Theoretical Schedules

An important fact that should be kept in mind is that a multicore processor is a synchronous system, because cores use the same clock, therefore transactions execute synchronously, in parallel, on workers' cores. The experiments conducted in [10-11] confirm this fact. We take this fact as a postulate when formalizing PSTM transaction scheduler architecture and deriving the theoretical schedules for the given number of workers and test workloads.

In this section we derive the theoretical schedules for the scheduler architecture with two workers, because the results in [12] were given for this case, and also because these simple schedules are easy to understand and analyze even by researchers who are not too familiar with this topic. The theoretical schedules for the three test workloads are shown in the figures Figure 1, Figure 2, and Figure 3, respectively.

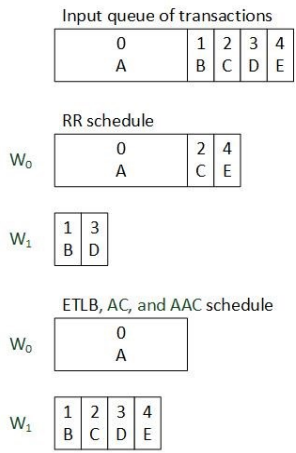

Figure 1: The theoretical schedules for CFW.

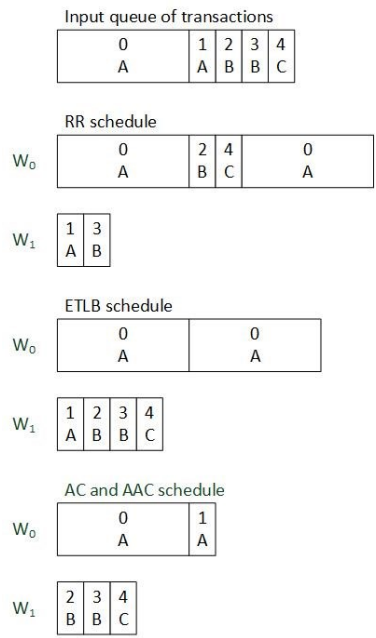

Figure 2: The theoretical schedules for CW-1.

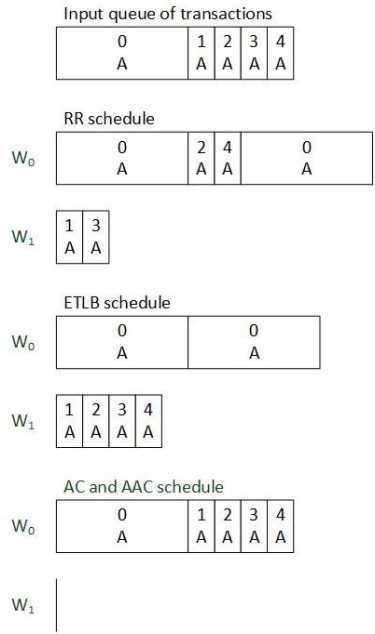

Figure 3: The theoretical schedules for CW-2.

In each figure, the queue with input transactions is shown at its top, where each transaction is labeled with its index (i.e. its ID in the CSP# model) and the name of the t-variable that it uses and updates (i.e. reads and writes to). The schedules for individual online transaction scheduling algorithms are shown below the input queue, where the two workers' queues are labeled with $W_0$ and $W_1$, respectively.

Figure 1 shows the expected schedule for the test workload CFW, which is conflict free, because all the transactions operate on different t-variables. Since there are two workers, RR algorithm works by modulo two, so in its first scheduling iteration it assigns even transactions to $W_0$ and odd to $W_1$. Because there are no conflicts, there are no transaction re-executions, and this initial schedule happens to be the complete schedule (i.e. the complete system evolution) with the makespan equal to 70, and the number of aborts equal to 0.

ETLB algorithm does better by balancing the load of the two workers, so it first assigns the transaction $T_0$ to $W_0$, and then it assigns all the remaining transactions to $W_1$. Since AC and AAC algorithms do the same when there are no conflicts, they produce the same complete schedule with the makespan equal to 50, and the number of aborts equal to 0.

Figure 2 shows the expected schedule for the test workload CW-1, which has two potential conflicts – the first conflict is between transactions $T_0$ and $T_1$ (because they operate on the t-variable A), and the second conflict is between $T_2$ and $T_3$ (over the t-variable B). Since both RR and ETLB algorithms are not aware of the existing conflicts, they produce the same initial schedules as in Figure1. However, in both of these initial schedules (for RR and ETLB algorithms) there is the conflict between $T_0$ and $T_1$ (over the t-variable A), and since $T_1$ finishes first, it gets committed, whereas $T_0$ gets aborted, rescheduled, and re-executed solo (i.e. without any other parallel transaction) in the second scheduling iteration. Therefore, RR produces the complete schedule with the makespan equal to 120 and the number of aborts equal to 1, whereas ETLB produces the complete schedule with the makespan 100 and the number of aborts equal to 1.

AC and AAC algorithms do better because they are aware of the existing conflicts. Since there are only two workers, AC does as good as AAC, and both algorithms in the first scheduling iteration make the conflict free schedule by first assigning: $T_0$ to $W_0$, $T_1$ again to $W_0$ (in order to avoid the potential conflict between $T_0$ and $T_1$), and $T_2$ to $T_4$ to $W_1$ (because by doing that they balance the load, i.e. minimize the makespan). So, both AC and AAC algorithms produce the complete schedule with the makespan 60, and the number of aborts 0.

Figure 3 shows the expected schedule for the test workload CW-2, which has potential conflicts among all the transactions (over the t-variable *A*). Since both RR and ETLB algorithms are not aware of the existing conflicts, they again produce the same initial schedules as in Figure 1 and Figure 2, but this time there are more conflicts than in Figure 2.

RR algorithm makes the initial schedule with the conflict between $T_0$ and $T_1$ and between $T_0$ and $T_3$, and since $T_1$ and $T_3$ finish before $T_0$, they get committed, whereas $T_0$ gets aborted, and re-executed solo in the second scheduling iteration. Alternatively, ETLB algorithm makes the initial schedule with the conflicts between $T_0$ and each of the transactions from $\{T_1, T_2, T_3, T_4\}$, and since the transactions $T_1$ to $T_4$ finish before $T_0$, they all get committed, whereas $T_0$ gets aborted, and re-executed solo in the second scheduling iteration. Therefore, RR produces the complete schedule with the makespan equal to 120 and the number of aborts equal to 1, whereas ETLB produces the complete schedule with the makespan 100 and the number of aborts equal to 1.

Both AC and AAC algorithms in the first scheduling iteration make the optimal conflict free schedule, which is therefore also the complete schedule, with the makespan equal to 90, and the number of aborts equal to 0.

Table 1 summarizes the expected results, for the makespan (*ms*) and the number of aborts (*na*), derived in this section together with the corresponding results from [12] (shown in the column Previous results). The abbreviation NA in Table 1 stands for not available. We test the results from [12] by comparing them with the expected results. Obviously, they are different (see Table 1). Therefore, we thoroughly examined the previous CSP# model from [12], and we report our findings in the next section.

Table 1: The expected results versus the previous results

| Load & Alg. | | Expected results | | Previous results | |
| --- | --- | --- | --- | --- | --- |
| Load | Alg. | *ms* | *na* | *ms* | *na* |
| CFW | RR | 70 | 0 | 1.0298s | 0 |
| | ETLB | 50 | 0 | 0.0321s | 0 |
| | AC | 50 | 0 | 0.0375s | 0 |
| | AAC | 50 | 0 | NA | NA |
| CW-1 | RR | 120 | 1 | 1.6344s | 2 |
| | ETLB | 100 | 1 | 1.2286s | 1 |
| | AC | 60 | 0 | 0.0137s | 0 |
| | AAC | 60 | 0 | NA | NA |
| CW-2 | RR | 120 | 1 | 2.6422s | 6 |
| | ETLB | 100 | 1 | 1.9130s | 7 |
| | AC | 90 | 0 | 0.1094s | 0 |
| | AAC | 90 | 0 | NA | NA |

### 2.3 Analysis Findings

In this section, we briefly explain the shortcomings that we discovered by detailed examination of the previous CSP# model from [12], see the summary in Table 2. Obviously the results for the makespan from [12] are different than the expected results (the former are neither equal nor proportional to the latter). Actually, the metric called Time Used (TU) reported by the model checker PAT was interpreted as equal to the makespan of the subject system under verification, and the consequence of this oversight is that the results for both the makespan and the throughput reported in [12] are not as expected.

Table 2: The summary of analysis findings

| Number | Finding |
|---|---|
| 1 | TU interpreted as the makespan |
| 2 | Shortcoming in the macro isConflict (the case $i = x$) |
| 3 | Pessimistic concurrency control |
| 4 | Asynchronous transactions' execution |
| 5 | The number of workers fixed to *Index* = 2 |
| 6 | AAC algorithm not supported |

The fact that the number of aborts from [12] are not as expected (see Table 1, the row for the CW-1 workload and RR algorithm, and the rows for the CW-2 workload and RR and ETLB algorithms) immediately indicated that the real schedules from [12] are different from the expected schedules derived in the section 2.2. A rather tedious reconstruction of the real schedules from the model checker PAT's log files confirmed this expectation. Finding the root causes for these behaviors proved to be a difficult and laborious task, because debugging CSP# models reduces to the analysis of PAT's log files or PAT's simulations.

The three main root causes of the discrepancies between the real and the expected schedules are discovered in the findings 2-4. Firstly, the CSP# macro isConflict from [12], which checks whether there are conflicts between the next transaction to be scheduled $x$ and any already scheduled transaction $i$, included the case $i = x$. Secondly, the CSP# model from [12] uses the pessimistic concurrency control, whereas the real PSTM uses the optimistic concurrency control (see the section 2.1 in [2] for the corresponding definitions). Thirdly, the workers within the CSP# model from [12] execute transactions asynchronously, and therefore this model violates the postulate of the synchronous transaction execution introduced in the section 2.2.

### 3 NEW CSP# MODEL

In this section we briefly present the new CSP# model, which evolved from the previous model [12]. The next subsections present the corrections and extensions that we made according to the findings 1-4 in the section 2.3 (the extensions for the findings 5 and 6 are skipped because of the space limits).

### 3.1 Makespan Calculation

Makespan (called span in [19]) is the longest time to execute the strands (i.e. instructions) along any path in the dag (directed acyclic graph), which represents a multithreaded computation [19]. Generally, makespan calculation for a given graph of computation (i.e. dag) is based on the two basic laws, which apply to the cases when the subgraphs of computation are connected serially or in parallel, see the section 27.1 in [19]. Here we define these laws in terms of CSP processes.

Let $T_\infty(P)$ and $T_\infty(Q)$ be the makespans of the processes *P* and *Q*, respectively. The following law gives the makespan for the sequential composition of *P* and *Q*:

$$T_\infty(P; Q) = T_\infty(P) + T_\infty(Q)$$

The following law gives the makespan for the parallel composition of *P* and *Q*:

$$T_\infty(P \parallel Q) = \max(T_\infty(P), T_\infty(Q))$$

where max is the maximum function.

Since the PSTM online transaction algorithms govern iterative execution of incoming transactions, see Figure 1 to 3, the complete makespan is the sum of makespans for individual iterations (by the law for the sequential composition). Similarly, within an iteration, the makespan of the series of transactions assigned to a worker is the sum of the durations of all the transactions in that series. Alternatively, the makespan for an iteration is the maximal makespan for the series of transactions executed in parallel by the available workers (by the law for the parallel composition).

We extended the processes representing the PSTM online transaction scheduling algorithms with new parts implementing the makespan calculation as explained above. Since the makespan calculation is the same for all the algorithms, we show this extension only for RR algorithm in Algorithm 1 (RR algorithm is modeled by the processes *Scheduler1* and *Scheduler1_1*).

ALGORITHM 1: The makespan calculation in RR algorithm

```
// The variable makespan is initially set to 0
1: Scheduler1() =
2:      comAS?READY -> Scheduler1_1();
3: Scheduler1_1() =
4:      if(QueueIn.Count() == 0) {
5:          atomic { upMkSp {
6:              call(findmax, load);
7:              makespan = makespan + load[Max];
8:              call(calcidleticks, 0); }->
9:              ...
```

When executing RR algorithm, the scheduler initially behaves as the process *Scheduler1*. After receiving the signal READY from the application process, it behaves as the process *Scheduler1_1* (line 2). Once it scheduled all the transactions from the queue *QueueIn* (line 4), it calls the macro findmax (line 6), which operates on the array *load*, whose elements contain loads (total execution times for the series of transactions) assigned to the individual workers. The macro findmax sets the variable *Max* to the index of the maximal load in the array *load*, which is equal to the maximal makespan in the current scheduling iteration. Finally, it calculates the *makespan* of the current iteration by adding *load*[*Max*] to the *makespan* of the previous iteration (line 7). Note: line 8 is not relevant here and will be explained later.

## 3.2 Fixed Macro isConflict

The fixed macro isConflict is shown in Algorithm 2. We added the condition ($i$ != $x$) in the line 9 in order to skip checking the conflict between the transaction $x$ and itself.

ALGORITHM 2: The fixed macro isConflict

```
01: var IsConflict = 0;
02: #define isConflict(x, t1, t2)
03: {
04:        var i = 0;
05:        IsConflict = 0;
06:        while (i < TNum && IsConflict != 1) {
07:                if( T_isScheduled[i] == 1) {
08:                        if((T_StartTime[i] < t2 && t2 <= T_EndTime[i]) ||
                              (t1 < T_EndTime[i] && T_EndTime[i] <= t2)) {
09:                                if((i != x) && (T_Var[i] == T_Var[x])) IsConflict=1;
10:                        }
11:                }
12:                i++
13:        }
14: };
```

## 3.3 PSTM's Optimistic Concurrency Control

We introduced the model extension shown in Algorithm 3, which models PSTM from the point of its optimistic concurrency control. This model is much simpler than the complete PSTM models developed in [13-14], but it serves its purpose, and it was intentionally made as such, in order to keep the model state space exploration fast and feasible.

ALGORITHM 3: The PSTM's optimistic concurrency control model

```
01: var T_VarVer[TNum]:{0..TNum-1} = [0, 0, 0, 0, 0];
02: Pstm() =
03:        worker2pstm?i.req.key.ver ->
04:        if(req == GetVars) {
05:                pstm2worker[i]!T_VarVer[key] -> Pstm()
06:        } else { // commitVars
07:                if(ver == T_VarVer[key]) {
08:                        {T_VarVer[key] = (T_VarVer[key]+1)%TNum} ->
                           pstm2worker[i]!1 -> Pstm()
09:                } else {
10:                        pstm2worker[i]!0 -> Pstm()
```

```
11:            }
12:       }
```

The PSTM dictionary is model by the array *T_VarVer* whose elements contain the versions of the respective t-variables. Since the total number of transactions within the given workloads (CFW, CW-1, CW-2) is bounded and equal to the constant TNum = 5, we decided to bound the number of versions of each t-variable by the same constant, in order to reduce the state space to be explored. The number of a t-variable versions is effectively bounded by counting them with modulo TNum, see the line 8.

The PSTM itself is modeled as the process *Pstm* (see line 2). All the messages from the worker processes are sent to the process *Pstm* over the channel *worker2pstm* (that corresponds to the real PSTM's input queue), whereas the replies from *Pstm* to the worker process *i* are sent over the channel *pstm2worker*[*i*].

The compound messages exchanged over the channel *worker2pstm* have the format *i.req.key.ver*, where *i* is the index (i.e. ID) of the worker, *req* is the type of the request (there are two types of requests that are encoded with the constants GetVars and CommitVars, which correspond to the PSTM API functions getVars and commitVars, respectively), *key* is the index (i.e. ID) of the t-variable, and *ver* is the version of the t-variable.

The replies transmitted over channels have a single element whose semantics depend on the type of the request. In case of the type GetVars, the reply is the t-variable' version, whereas in the case of the type CommitVars, the reply is either 0 or 1 whether the transaction gets aborted or successfully committed, respectively.

After receiving the message *i.req.key.ver*, the process *Pstm* checks the type of the request (line 3). In case of the request type GetVars (line 4), it returns the current version of the t-variable with index key (line 5). Alternatively, in case of the request type CommitVars, *Pstm* checks whether the version *ver* from the input message is equal to the current version of the t-variable with the index *key* (line 7). If these versions are equal, *Pstm* increments (with the modulo TNum) the current version of the t-variable with the index *key* and sends the reply 1 (signaling successful commit), otherwise it sends the reply 0 (signaling abort).

*Pstm* provides a mechanism for optimistic concurrency control by servicing two types of requests made by (synchronous) concurrent workers executing transactions. Each transaction starts with the GetVars request (to get the current version of the target t-variable) and ends with the CommitVars request (to perform the commit operation and get its result).

### 3.4 Synchronous Transaction's Execution

We changed and extended the processes modeling the behavior of workers as shown in Algorithm 4 (unimportant parts are skipped). The worker *i* initially behaves as the process *Worker*(*i*) (line 1). After receiving the signal READY from the scheduler, it behaves as the process *Worker_1*(*i*) (line 2).

ALGORITHM 4: The workers' behavior

```
01: Worker(i) =
02:       comSW[i]?READY->Worker_1(i);
03: Worker_1(i) =
04:       if(Queue[i].Count() == 0) {
05:               Worker_2(i); output!done -> Worker(i)
```

```
06:        } else {
07:              Working(i);
                 ...
08: Worker_2(i) =
09:       if(idleTicks[i] > 0) {
10:              tick -> tau{idleTicks[i]--} -> Worker_2(i)
11:       } else {Skip};
12: Working(i) =
13:       tick ->
14:       worker2pstm!i.GetVars.currentT_Var[i].0 ->
15:       pstm2worker[i]?tvarver ->
16:       {currentT_VarVer[i] = tvarver; workertime[i]++} ->
17:       Working_1(i);
18: Working_1(i) =
19:       tick ->
20:       if(workertime[i] < currentT_Time[i]) {
21:              working{workertime[i]++} ->
22:              Working_1(i)
23:       } else {
24:              worker2pstm!i.CommitVars.currentT_Var[i].
                 currentT_VarVer[i] ->
25:              pstm2worker[i]?resp ->
26:              {currentT_Cmt[i] = resp; workertime[i] = 0} ->
27:              Skip
28:       };
```

The process *Worker_1(i)* (line 3) workers iteratively. In each iteration it checks whether there is a transaction to be processed in its input queue *Queue[i]* (line 4). If yes, the process dequeues the transaction, estimates the duration of the transaction by calling the macro estimateTime, sets the variables related to the transaction (not shown in Algorithm 4), and continues behaving as the process *Working(i)* (line 7). Otherwise, if there are no (more) transactions to be processed by the *Worker_1(i)*, it transforms to the process *Worker_2(i)* (line 5).

By the definition of the parallel composition operator ||, all the parallel processes must simultaneously perform the common events from their alphabets. Since the event *tick* is common for all the workers, they synchronize using the so called lock-step synchronization, i.e. they engage in the event *tick* simultaneously. This means that all the workers must perform the same number of ticks (i.e. tick events), *nt*, per scheduling iteration. This number is easy to calculate. Let $nt_i$ be the total load plus the number of transactions allocated to the worker *i* (we add the number of transactions to the worker, because starting each transaction requires one *tick*). Then, *nt* is the maximal $nt_i$, $nt_m$ ($nt_i$ for the worker *i* = *m*), for *i* = 0, …, Index – 1. More precisely $nt_i$ is defined as:

$$nt_i = load_i + \text{count}(Queue_i)$$

and $nt_m$ is defined as:

$$nt = nt_m = \max_i nt_i$$

where $\max_i nt_i$ is the maximum of all the $nt_i$.

Further, let $it_i$ be the number of idle ticks that should be executed by the worker $i$ after it processed all the transactions form its input queue, $Queue_i$:

$$it_i = nt - nt_i$$

where $nt$ and $nt_i$ are as defined above.

The processes *Working*(*i*) and *Working_1*(*i*) model the behavior of the worker processing a transaction. More precisely, the process *Working*(*i*) models the start of the transaction, whereas the process *Working_1*(*i*) models the rest of the transaction.

The process *Working*(*i*) (line 12) does the following steps: (i) send the GetVars request for its t-variable to the process *Pstm* (line 14) and receive the current version of this t-variable in the reply from *Pstm* (line 15), and (ii) store the current version of its t-variable and increment its working time by updating the corresponding model variables (line 16).

The process *Working_1*(*i*) (line 18) checks whether it has to do more processing, and if yes, it increments its working time (line 21). Otherwise, it reached the end of the transaction, and it performs the following steps: (i) sent the CommitVars request to the process *Pstm* (line 24) and receive the corresponding reply in the variable *resp*, which is 1 if the transaction got successfully committed, otherwise 0 (line 25), and (ii) store the reply and reset the working time (for the next scheduling iteration) by updating the corresponding model variables (line 26).

After the worker *i* processed all the transactions from its input queue it transforms to the process *Worker_2*(*i*) (line 8), which simply checks whether the number of its idle ticks ($it_i$) is (still) greater than zero (line 9), and if yes, decrements it (line 10); otherwise it completes its performance (line 11).

## 4  FORMAL VERIFICATION

This section presents the formal verification based on the new CSP# model. The next two subsections present the verification results and the performance analysis, respectively.

### 4.1  Verification Results

We first introduce and briefly explain the conditions that are used in the assertions that were automatically checked by the model checker PAT. Let *snum* be the number of the successfully executed transactions (also recall that *na* and *ms* are the number of aborts and the makespan, respectively). Then we define the conditions as follows. The condition *Done* requires that *snum* is equal to TNum, which means that all the transactions have been successfully executed. The condition *MaxNA* requires that *na* is nonnegative and that Done is satisfied. Similarly, the condition *MaxMS* requires that *ms* is nonnegative and that *Done* is satisfied.

Next we briefly explain the five assertions that were checked for each version of the system, where the version of the system is defined by the given number of workers and the given algorithm. The first four assertions are essentially the same as in [12], whereas the fifth is the new one. The first one corresponds to a safety property, and the rest correspond to liveness properties.

Assertion 1: Claims that the system is deadlock free.

Assertion 2: Claims that the system reaches a state satisfying the condition *Done*.

Assertion 3: Claims that all the system's evolution paths satisfy the CSP# LTL formula []<>*comSA*.complete, which means that always ([]) eventually (<>) the signal complete is sent from the scheduler to the application, over the channel *comSA*.

Assertion 4: Claims that the system reaches a state satisfying the condition MaxNA over a path that maximizes *na*, and reports that maximal value of *na* (using the assertion clause "with").

Assertion 5: Claims that the system reaches a state satisfying the condition MaxMS over a path that maximizes the expression (*ms* + *na*), and reports that maximal value of that expression.

By using PAT, we checked these five assertions for the system versions with two, three, and four workers (i.e. Index = 2, 3, 4), and four each of the four PSTM online scheduling algorithms (RR, ETLB, AC, and AAC), i.e. in total for twelve system versions (3 x 4 = 12). All the sixty assertions (12 x 5 = 60) were found to be valid (i.e. satisfied).

The third assertion for the case with four workers, CW-1 workload, and RR algorithm was the most time consuming to validate. The verification statistics reported by PAT for this case is the following: 6738583 visited states, 35068677 passed transitions, 219s of used time, and 7361128.96KB of used memory (on a PC with 16GB DDR4 memory and CPU i7-8750H 2.2 GHz with turbo bust to 4.1 GHz).

Additionally, we manually checked the values for *na* and *ms* reported by the last two assertions, and they matched the expected results. The expected results for *na* and *ms* and the two workers (Index=2) are already given in Table 1, whereas the expected results for the cases with the three and four workers are given in Table 3 (because of space limits we skipped their derivation).

Table 3: The expected results for three and four workers

| Load & Alg. | | Three workers | | Four workers | |
| --- | --- | --- | --- | --- | --- |
| Load | Alg. | *ms* | *na* | *ms* | *na* |
| CFW | RR | 60 | 0 | 60 | 0 |
| | ETLB | 50 | 0 | 50 | 0 |
| | AC | 50 | 0 | 50 | 0 |
| | AAC | 50 | 0 | 50 | 0 |
| CW-1 | RR | 110 | 1 | 110 | 2 |
| | ETLB | 100 | 1 | 100 | 2 |
| | AC | 70 | 0 | 70 | 0 |
| | AAC | 60 | 0 | 60 | 0 |
| CW-2 | RR | 160 | 3 | 210 | 6 |
| | ETLB | 200 | 6 | 200 | 6 |
| | AC | 90 | 0 | 90 | 0 |
| | AAC | 90 | 0 | 90 | 0 |

### 4.2 Performance Analysis

In this section, we analyze the performance of the PSTM online transaction scheduling algorithms from the perspectives of the makespan and the system throughput.

We start by introducing the notion of the number of independent transactions within a workload, which we use to quantitatively characterize the level of parallelism for a given workload, *L*.

***Definition 1.*** *The number of independent transactions, nit, is the max number of transactions from L that could be scheduled online (i.e. without changing the transactions arrival order), in parallel, on an infinite number of processors, without a conflict.*

It can easily be seen that the values of *nit* for CW-2, CW-1, and CFW, are 1, 3, and 5, respectively. So, CW-2 has the lowest level of parallelism, and CFW has the highest.

Next we constructed Table 4 by calculating the average makespan (using Tables 1 and 3) for each algorithm and each workload (which is characterized by its *nit* value). The values of the average *ms* for RR, ETLB, AC, and AAC algorithms are denoted as *ms-rr*, *ms-etlb*, *ms-ac*, and *ms-aac*, respectively. The values, from Table 4, for individual algorithms are illustrated with the corresponding curves in Figure 4.

Table 4: The makespan vs the level of parallelism

| nit | ms-rr | ms-etlb | ms-ac | ms-aac |
|---|---|---|---|---|
| 0 | 163.33 | 166.66 | 90 | 90 |
| 3 | 113.33 | 100 | 66.66 | 60 |
| 5 | 63.33 | 50 | 50 | 50 |

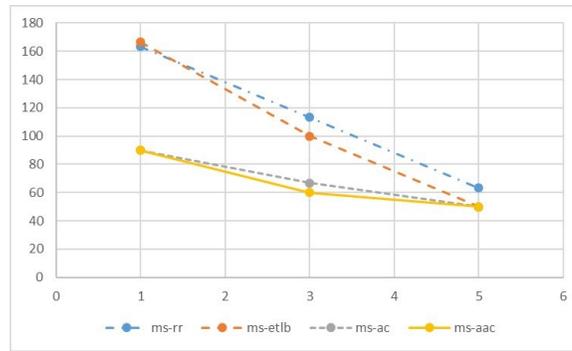

Figure 4: The makespan vs the level of parallelism.

Further on, the system throughput is defined as the ratio of the number of transactions in a workload and the makespan [12]. In this paper, in order to get more realistic values for throughput, we assume that makespan values in Tables 1 and 3 are in milliseconds, because durations of transactions in the experiments in [10-11] where in order of tens of milliseconds.

Next we constructed Table 5 by calculating the average throughput (using Table 4) for each algorithm and each workload (the number of transactions in each workload is five). The values of the average throughput for RR, ETLB, AC, and AAC algorithms are denoted as *th-rr*, *th-etlb*, *th-ac*, and *th-aac*, respectively. The values, from Table 5, for individual algorithms are illustrated with the corresponding curves in Figure 5.

Table 5: The throughput vs the level of parallelism

| nit | th-rr | th-etlb | th-ac | th-aac |
|---|---|---|---|---|
| 0 | 32.33 | 33.33 | 55 | 55 |
| 3 | 44 | 50 | 75 | 83 |
| 5 | 79 | 100 | 100 | 100 |

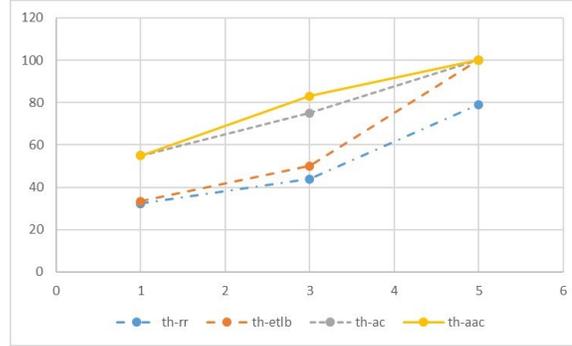

Figure 5: The throughput vs the level of parallelism.

Finally, both the data in Tables 4 and 5, and the shape of the curves in Figure 4 and Figure 5, confirm that according to their performance, both in terms of the makespan and the throughput, the PSTM online transaction scheduling algorithms are ranked as follows: (i) AAC is the best, (ii) AC is worse than AAC, (iii) ETLB is worse than AC, and (iv) RR is the worse. This finding is well aligned with the experimental and theoretical evaluations made in the previous research in [10-11].

## 5 CONCLUSIONS

Contemporary society more and more needs a trustworthy software, and therefore software verification is of paramount importance. An important lesson learned from this research is that using formal methods, e.g. CSP, in isolation is not sufficient. As a solution we demonstrated an approach to the complete formal verification of trustworthy software by jointly using both formal verification and it's testing.

In the paper, we applied this approach and conducted the complete verification of PSTM online transaction scheduler and accompanying scheduling algorithms, through the following steps: (i) we derived the expected results (schedules and their makespan and the number of aborts), (ii) we analyzed the previous CSP model by testing the model checker PAT results and manual inspection of the relevant parts of the model, (iii) according to the results of this analysis, we corrected and extended the CSP model, and (iv) we analyzed the performance of PSTM online transaction scheduling algorithms from the perspective of makespan, number of aborts, and throughput, and got the results that were as expected and well aligned with the previous research [10-11].

The main limitations of this paper are: (i) the verification is based on the three particular workloads, and (ii) each transaction in these workloads operates on a single t-variable. In our future work, we plan to make further extensions in the CSP model, which are required for the test workloads from the previous research [10-11].

## ACKNOWLEDGMENTS

This work was partially supported by the: Ministry of Education, Science and Technology Development of Republic of Serbia under Grant no: 451-03-68/2020-14/200156.

## REFERENCES

[1]  M. Herlihy and J. E. B. Moss. 1993. Transactional memory: Architectural support for lock-free data structures. In Proceedings of the 20th Annual International Symposium on Computer Architecture. ACM, New York, NY, 289-300. https://doi.org/10.1145/165123.165164

[2]  T. Harris, J. R. Larus, and R. Rajwar. 2010. Transactional Memory, 2nd edition, Morgan and Claypool.

[3]  N. Shavit and D. Touitou. 1995. Software transactional memory. In Proceedings of the 14th Annual ACM Symposium on Principles of Distributed Computing. ACM, New York, NY, 204-213. https://doi.org/10.1145/224964.224987

[4]  R. M. Yoo and Hsien-Hsin S. Lee. 2008. Adaptive transaction scheduling for transactional memory systems. In Proceedings of the 12th annual symposium on Parallelism in algorithms and architectures. ACM, New York, NY, 169–178. https://doi.org/10.1145/1378533.1378564

[5]  M. Ansari, M. Luján, C. Kotselidis, K. Jarvis, C. Kirkham, and I. Watson. 2009. Steal-on-abort: Improving transactional memory performance through dynamic transaction reordering. Lecture Notes in Computer Science, vol 5409. Springer, Berlin, Heidelberg. https://doi.org/10.1007/978-3-540-92990-1_3

[6]  S. Dolev, D. Hendler, and A. Suissa. 2008. CAR-STM: scheduling based collision avoidance and resolution for software transactional memory. In Proceedings of the 27th ACM symposium on Principles of distributed computing. ACM, New York, NY, 125–134. https://doi.org/10.1145/1400751.1400769

[7]  M. Popovic and B. Kordic. 2014. PSTM: Python software transactional memory. In Proceedings of the 22nd Telecommunications Forum. IEEE, 1106-1109. https://doi.org/10.1109/TELFOR.2014.7034600

[8]  M. Goldstein, E. Fredj, B. Gerber. 2011. A New Hybrid Algorithm for Finding the Lowest Minima of Potential Surfaces: Approach and Application to Peptides. Journal of Computational Chemistry 32, 1785–1800. https://doi.org/10.1002/jcc.21755

[9]  B. Kordic, M. Popovic, M. Popovic, M. Goldstein, M. Amitay, and D. Dayan. 2019. A Protein Structure Prediction Program Architecture Based on a Software Transactional Memory. In Proceedings of the 6th Conference on the Engineering of Computer Based Systems. ACM, Article 1, 9 pages. https://doi.org/10.1145/3352700.3352701

[10] M. Popovic, B. Kordic, and I. Basicevic. 2017. Transaction Scheduling for Software Transactional Memory. In Proceedings of the 2nd IEEE International Conference on Cloud Computing and Big Data Analysis. IEEE, 191-195. https://doi.org/10.1109/ICCCBDA.2017.7951909

[11] M. Popovic, B. Kordic, M. Popovic, I. Basicevic. 2019. Online Algorithms for Scheduling Transactions on Python Software Transactional Memory. Serbian Journal of Electrical Engineering 16, 1, 85-104. https://doi.org/10.2298/SJEE1901085P

[12] C. Xu, X. Wu, H. Zhu, M. Popovic. 2019. Modeling and Verifying Transaction Scheduling for Software Transactional Memory using CSP. In Proceedings of the 13th Theoretical Aspects of Software Engineering Symposium. IEEE, 240-247. https://doi.org/10.1109/TASE.2019.00009

[13] B. Kordic, M. Popovic, S. Ghilezan. 2019. Formal Verification of Python Software Transactional Memory Based on Timed Automata. Acta Polytechnica Hungarica 16, 7, 197-216. https://doi.org/10.12700/APH.16.7.2019.7.12.

[14] A. Liu, H. Zhu, M. Popovic, S. Xiang, L. Zhang. 2020. Formal Analysis and Verification of the PSTM Architecture Using CSP. Journal of Systems and Software 165, 1–14. https://doi.org/10.1016/j.jss.2020.110559

[15] C.A.R. Hoare. 1985. Communicating Sequential Processes. Prentice/Hall International.

[16] Y. Si, J. Sun, Y. Liu, J. S. Dong, J. Pang, S. J. Zhang, and X. Yang. 2014. Model checking with fairness assumptions using PAT. Frontiers of Computer Science 8, 1, 1–16. https://doi.org/10.1007/s11704-013-3091-5

[17] E. Koskinen, M. Parkinson. 2015. The Push/Pull Model of Transactions. In Proceedings of the 36th ACM SIGPLAN Conference on Programming Language Design and Implementation. ACM, 186-195. https://doi.org/10.1145/2737924.2737995

[18] M. Popovic, M. Popovic, S. Ghilezan, B. Kordic. 2019. Formal Verification of Python Software Transactional Memory Serializability Based on the Push/Pull Semantic Model. In Proceedings of the 6th Conference on the Engineering of Computer Based Systems. ACM, Article 6, 8 pages. https://doi.org/10.1145/3352700.3352706

[19] T.H. Cormen, C.E. Leiserson, R.L. Rivest, and C. Stein. 2009. Introduction to Algorithms, 3rd Edition. The MIT Press.

[20] R. Alur, D. L. Dill. 1994. A theory of timed automata. Theoretical Computer Science 126, 2, 183-235. https://doi.org/10.1016/0304-3975(94)90010-8

[21] G. Behrmann, A. David, and K. G. Larsen. 2004. A Tutorial on Uppaal. Lecture Notes in Computer Science, vol 3185. Springer, Berlin, Heidelberg. https://doi.org/10.1007/978-3-540-30080-9_7

[22] B. Berard, J. M. Bidoit, A. Finkel, F. Laroussinie, A. Petit, L. Petrucci, Ph. Schnoebelen, and P. McKenzie. 1999. Systems and Software Verification. Springer.

[23] Testing. Cambridge online dictionary. Retrieved March 16, 2021 from https://dictionary.cambridge.org/dictionary/english/testing